\documentclass[11pt]{article}
\usepackage{graphicx}

\def\SDiff{S\kern-1.5pt di \kern-1pt f \kern-1.5pt f}
\def\ltsima{$\; \buildrel < \over \sim \;$}
\def\simlt{\lower.5ex\hbox{\ltsima}}
\def\gtsima{$\; \buildrel > \over \sim \;$}
\def\simgt{\lower.5ex\hbox{\gtsima}}

\begin{document}

\begin{center}

{\bf Ellipticity in Cosmic Microwave Background as a Tracer of Large-Scale Universe}

\end{center}

\vspace{0.2in}

\noindent V.G.Gurzadyan$^{1,2}$, 
C.L.Bianco$^2$, A.L.Kashin$^1$, H.Kuloghlian$^1$,
G.Yegorian$^{1,2}$

\vspace{0.2in}

$^1$ Yerevan Physics Institute, Yerevan, Armenia; $^2$ ICRANet, ICRA, Dipartimento
di Fisica, Universita La Sapienza, Roma, Italy 
\vspace{0.2in}

{\bf Abstract} - Wilkinson Microwave Anisotropy Probe (WMAP) 3-year data confirm the ellipticity 
of anisotropies of Cosmic Microwave Background (CMB) maps, 
found previously for Boomerang and WMAP 1-year high sensitivity maps. 
The low noise level of the WMAP latter data enable also to show 
that, the ellipticity is a property not described by the conventional 
cosmological model fitting the power spectrum of CMB. As a large scale anomaly, 
the ellipticity characteristics are consistent with the
effect of geodesics mixing occurring in hyperbolic Universe.
Its relation to other large scale effects, i.e. to suppressed low multipoles, 
as well as to dark energy if the latter is due to vacuum fluctuations, 
is then an arising issue. 

\vspace{0.2in}

The analysis of high signal to noise ratio 150 GHz Cosmic Microwave Background (CMB) radiation maps 
obtained in the Boomerang 1998 experiment \cite{Bern1},
\cite{Bern2} has recently revealed ellipticity about 2 of the temperature anisotropies \cite{ellipse}. 
Wilkinson Microwave Anisotropy Probe's (WMAP) 1-year data \cite{Ben03} confirmed the results of Boomerang \cite{WMAP1year}. 
Initially a signature of ellipticity was found for COBE-DMR maps \cite{GT}.
The effect was shown to exist 
at scales both larger and smaller than the horizon scale at the last scattering epoch
\cite{Gurz1}. 

There are various unexpected features detected in the CMB sky, such as the quadrupole-octopole correlation,
planarity of the octopole, the cold spot, the Axis of Evil, etc. (e.g. \cite{Erik,Copi1,Sch,Land,Copi2,Mag,Cr,Mc}),
which are being reanalysed by WMAP 3-year data \cite{Sp,Page,Hin}.  

WMAP 3-year data provide also a possibility to follow the ellipticity effect with
higher signal-to-noise ratio, thus checking not only the value
of the ellipticity but also the role of the noise in comparison with simulated maps.

Originally, the interest to the ellipticity was motivated by the effect 
of geodesic mixing which has to exist in hyperbolic spaces 
due to exponential divergence of close null geodesics in (3+1)-space \cite{GK1}. 
These properties of hyperbolic spaces are well known both in
geometry \cite{Pen} and in theory of hyperbolic dynamical
systems \cite{Anosov}.
Another descriptor for CMB maps, the Kolmogorov complexity, was suggested in \cite{G}.

The algorithms of study of the ellipticties in pixelized CMB maps
have been described in previous papers \cite{ellipse,WMAP1year} and are reduced,
first, to the determination of the anisotropy spots (excursion sets) at given temperature
thresholds, and then, to the definition of their semi-axes. Although motivated by the
geodesic mixing, the measurement of mean elongation
actually implies a more general content, namely, estimation of a Lyapunov exponent 
of any dynamical system which might be its reason.  

WMAP 3-year data not only confirmed the value of the ellipticity detected before
in CMB maps, but enabled to draw new conclusions. 
In previous papers we had mentioned the impossibility
of direct tracing of the effect via maps simulated for the power spectrum parameters. 
The reason is that a possible fit of simulations to real data could not be considered as
a proof of the model, since the ellipticity initially was not within the parameters of the
model, and hence any other model may fit better. 
However, now due to low noise level of WMAP 3-year data it appears possible to reveal
the practically genuine ellipticity and compare the sequence of data maps of various
noise level with simulated maps.    

We found that, first, for simulated signal maps (no noise)
the ellipticity is increasing for the large spots while for the data maps it is the contrary, decreasing. 
Second, the contribution of the WMAP 3-year noise in the simulated maps is not only far larger, 
but is opposite for spots of various size. Namely,
for more than 20-pixel spots the noise is increasing the ellipticity, while for more than 50 or 100-pixel spots
we see the contrary, a decrease. So, both, by behavior and amplitude,
the ellipticity of the simulated maps contradicts the real maps.

This implies that the ellipticity is an effect not included in the
cosmological model fitting the power spectrum.      
Note, that for $\Lambda CDM$ model the Boomerang power spectra simulated maps have 
ellipticity around 1.8 \cite{ellipse} for the noise level of
Boomerang, while for the once popular flat CDM model, with no noise, the mean
ellipticity is around 1.4 \cite{Bond}.

The independence of the effect  on the scale of the horizon 
at the last scattering surface shows that it cannot be related to the physical conditions there,
and hence, most probably we deal with a large scale effect. 
Note, non-zero curvature or non-trivial topology models (e.g. \cite{Efs03,Lum03,Aur04}) are among the models
proposed to explain the low quadrupole.

We used the WMAP 3-year 94 GHz (3.2mm) maps, i.e. the same channel as we had used for
the 1-year data \cite{WMAP1year}, 
since they have highest angular resolution (beam size 0.22$^o$ FWHM) and are 
least influenced by the Galactic synchrotron radiation. 

To study the role of the noise via the real data we produced a sequence of maps 
of various noise level
combining the data from four independent detectors $w_i$
into two independent channels $A =(w_1+w_2)/2$ and $B =
(w_3+w_4)/2$ in Healpix \cite{Gorski} representation with nside=512
(6.9 arcmin pixel side), when the pixels with galactic
latitude $|b|<20^{\circ}$ were excluded, as usual. 
The algorithms for the definition of the excursion sets on the sphere, as well
as for the definition of the two semi-axes and of ellipticity as of their ratio,
are described in in details in \cite{ellipse, Gurz1}. We outline that,
3 types of possible errors have been studied separately. First, the statistical
ones appear after avareging over the number of the spots for the given threshold, 
second, the bias of the
algorithms have been thoroughly studied via variation of the input parameters and 
algorithmic steps,
third, the systematic errors have been studied via the sequence of real maps
with different noise level.   

In Fig.1 we represent the average ellipticity of the anisotropy spots
counting more than 20, 50 and 100 pixels,
as a function of the temperature threshold for the $A+B$ map. 
The error bars in Fig.1 are statistical only, as they are
computed from the standard deviation of the ellipticities of all
the areas at a given threshold.
As we have shown in \cite{ellipse}, for areas containing
more than several tens of pixels, the algorithm bias does not
exceed 0.1 in the ellipticity.
The statistics of the spots is given in Fig. 2 and is quite 
consistent with WMAP 1-year data \cite{WMAP1year}. Fig. 3 exhibits
the frequency dependence of the ellipticity.  As compared to w band,
the ellipticity homogeneity is distorted, especially,
at 41 GHz (q band), where the Galactic foregroud is stronger (possibly, due to anomalous
dust emission with steep fall at 60 GHz) \cite{Hin,DL}. 
In Fig. 4 we compare two maps, without $|b|<20^{\circ}$ pixels (as
in Fig. 1) and without $|b|<30^{\circ}$, thus showing the absence of any essential role
of the Galactic disk. Fig. 5 shows the ellipticity in first and second yearly maps,
along with the 3-year 94 GHz one. 

The procedure first adopted in \cite{WMAP1year} to reveal the role of
the noise in the ellipticity using not the simulations but the real data, 
is repeated here.
Namely, we obtained new maps by adding the difference data $A-B$, which
contain only noise, to different combinations of the $w_i$ data,
which contain signal and noise. This procedure effectively
changes the signal to noise ratio (S/N) of the map. We analyzed
the following maps: $(A+B)/2$; $(A'+B)/2$, where $A' =
(w_1+A-B+w_2)/2$; $w_1$; $w_1+A-B$. As a result, we obtain (for details see \cite{WMAP1year})
a sequence of 4 maps with different normalized N/S ratio: for
WMAP 1-year maps the S/N yields ($0.5$; $0.56$; $1.0$; $1.71$), respectively,
and smaller by a factor $\sqrt 3$ for 3-year ones. 
So, now we get maps based on WMAP 3-year data, in addition to those
of WMAP 1-year ones, with different noise levels. The ellipticities
in each of them have been estimated and are given in Fig. 6.
We see that the behavior shown in Fig. 5 of \cite{WMAP1year}
is confirmed and continued for smaller noise-to signal values, 
namely, the noise is affecting more strongly the smaller
spots, along with clear monotonic convergence of ellipticities for all three
pixel count spots.

Could one confirm this monotonic behavior of the noise in the simulated maps?
To see this, we simulated CMB maps corresponding to the WMAP power spectrum without
noise and then with WMAP's 3-year noise level. The simulations of the maps have 
been preformed by standard procedure. First, the Healpix \cite{Gorski} software
was used to produce the signal no-noise maps corresponding the power spectrum for the 
cosmological parameters of WMAP3 data. Then, the noise was superposed
using WMAP available $w$ band values for $\sigma$ (the average for the band,
as well as for each four $w_i$, separately), weighted by $N_{obs}$, the
effective number of measurements (for details see http://lambda.gsfc.nasa.gov).  
The results for 50 simulated maps are given 
in Fig. 7. First, we see that, without noise the ellipticity is
increasing for larger spots, which is contrary to the behavior 
in WMAP maps, as shown in Fig. 6. Second, the role of the
noise is opposite for more than 20 and 50,100-pixel spots, which is
strongly contradicting the monotonic behavior in real maps (Fig.6).
Moreover, even for 20-pixel spots when simulated maps show the noise increasing 
the ellipticity  as in real maps, numerically it implies no noise
ellipticity about 2, which is inconsistent with Fig. 6; with high confidence level
of incompatibility, up to 10$\sigma$, in the values of the ellipticities.   

Thus, the prediction in \cite{WMAP1year} that,
both independence or strong dependence of the
results on the noise in the simulated maps can be equally misleading, is confirmed.
We now found, that the behavior of the noise level dependence is, moreover,
not compatible with properties of the real maps.

WMAP 3-year data confirm the ellipticity around 2 in the CMB sky,
found before for COBE, Boomerang and WMAP's 1-year maps.

In addition, the low noise level of WMAP 3-year data enable to
conclude that, the ellipticity is a property
not described by the cosmological model fitting the power spectrum
of CMB. 
This is shown via the revealed incompatibility in the behavior of the noise, 
as well as in the dependence of the ellipticity of the size of the spots 
in the WMAP's and simulated maps. Namely, for WMAP's data a clear monotonic
behavior of the noise in the ellipticity is shown.   
While in the simulated maps, for more than 50 and 100 pixel
spots the role of the noise is contrary to that
in the real maps. For more than 20-pixel spots, even though the simulated behavior is 
compatible with real data, numerically there is 10$\sigma$ discrepancy between 
the simulated and measured ellipticities.
Finally, simulated maps violate the monotonic noise behavior in the real data.

If the ellipticity is due to geodesic mixing (with properties \cite{GK1}
fully corresponding to the measured ones)
or to another large scale effect, then
the interesting question is, how it is linked with the low quadrupole of the CMB \cite{Copi}, 
as well as with the dark energy, if the
latter is due to relevant vacuum fluctuations defined by the boundary
conditions of the Universe \cite{GX,V,DG}.

We are thankful to P. de Bernardis and his team for help and numerous discussions; G.Polenta
is particularly thanked for help with simulations of CMB maps.

\bigskip
\bigskip

\begin{figure}[htp]
\begin{center}
\includegraphics[height=8cm,width=12cm]{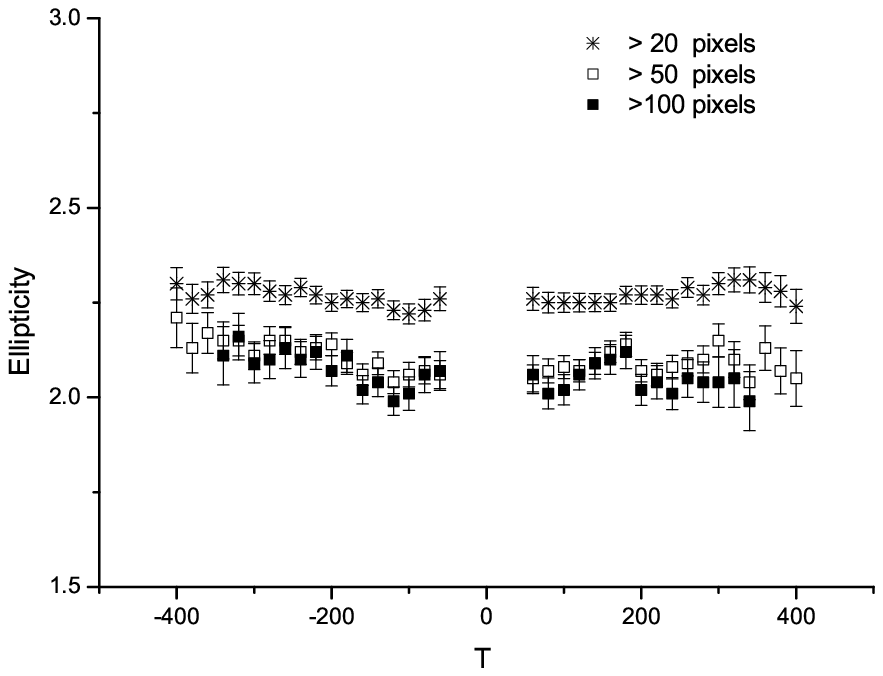}
\caption{Ellipticity of anisotropies vs temperature 
threshold (in $\mu K$)  for spots containing more than 20, 50 and 100 pixels in
WMAP 3-year 94 GHz sum map A = (w1+w2)/2 and B = (w3+w4)/2. The
error bars are statistical only.}
\end{center}
\end{figure}

\begin{figure}[htp]
\begin{center}
\includegraphics[height=8cm,width=12cm]{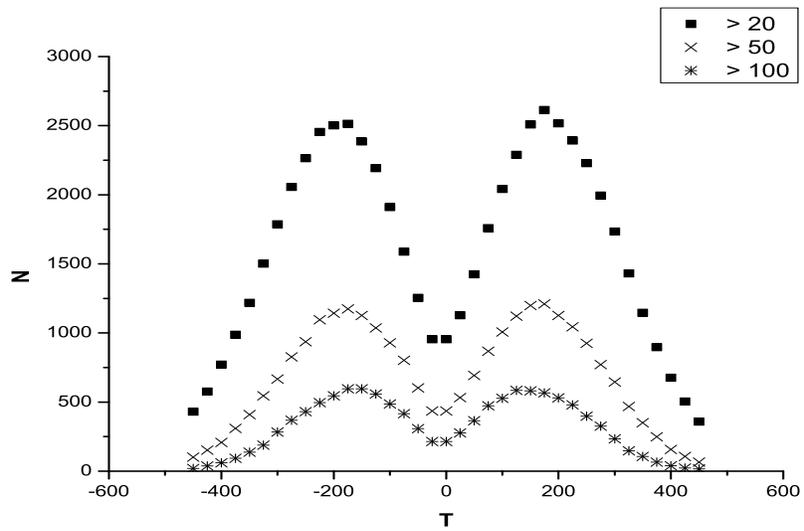}
\caption{The number of anisotropy spots containing more than 20, 50 and 100
pixels vs the temperature threshold; cf. with WMAP 1-year data in \cite{WMAP1year}}
\end{center}
\end{figure}

\begin{figure}[htp]
\begin{center}
\includegraphics[height=8cm,width=12cm]{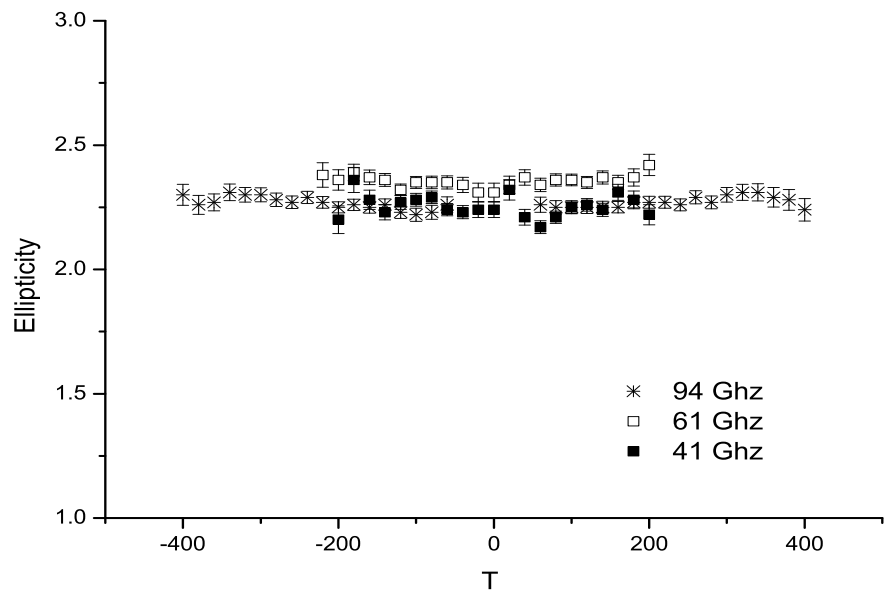}
\caption{Band dependence of the ellipticity; this and Figs.4,5 are for more than 20 pixel spots.}
\end{center}
\end{figure}

\begin{figure}[htp]
\begin{center}
\includegraphics[height=8cm,width=12cm]{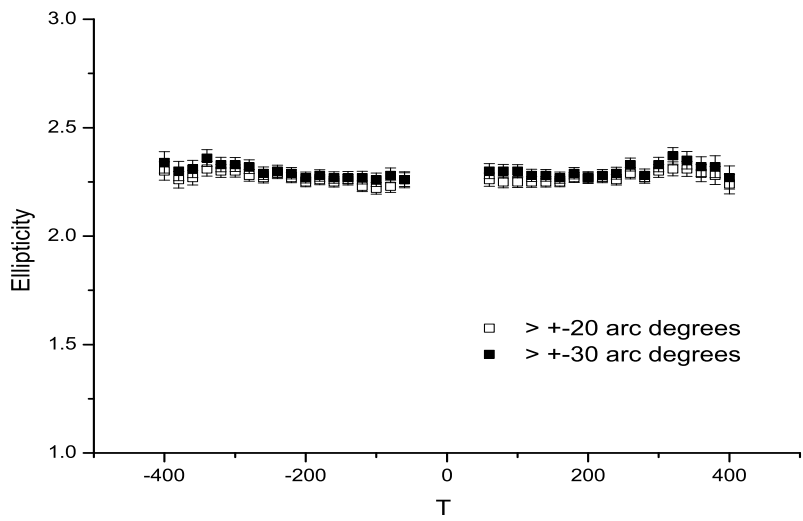}
\caption{Ellipticity when 
$|b|<20^{\circ}$ and $|b|<30^{\circ}$ pixels are excluded.}
\end{center}
\end{figure}

\begin{figure}[htp]
\begin{center}
\includegraphics[height=8cm,width=12cm]{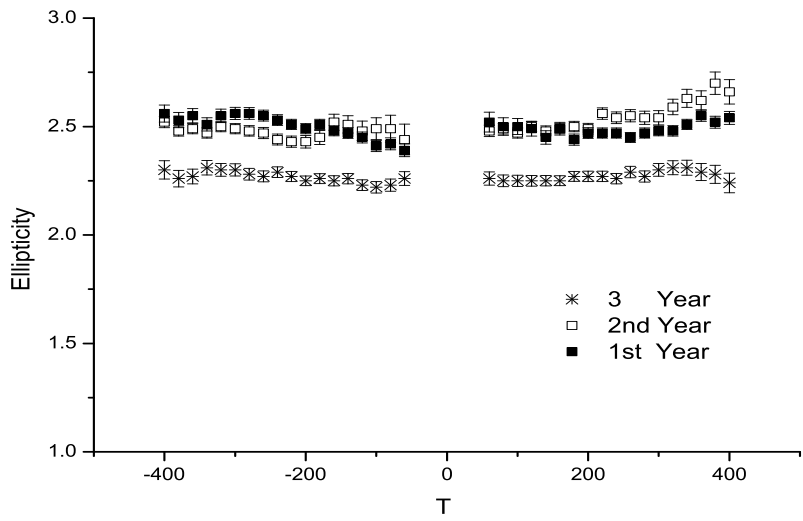}
\caption{Ellipticity in 1st, 2nd yearly, and 3-year (i.e. the sum of three years) maps.}
\end{center}
\end{figure}

\begin{figure}[htp]
\begin{center}
\includegraphics[height=8cm,width=12cm]{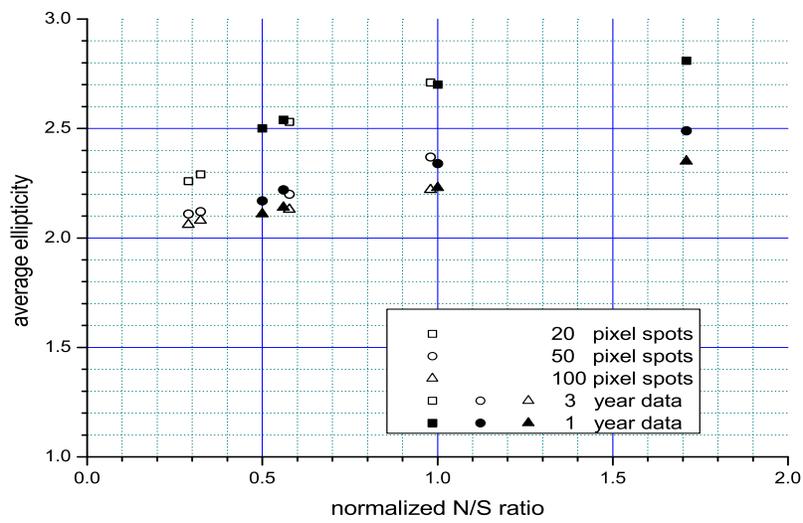}
\caption{Average ellipticity vs noise-to-signal ratio for WMAP 1-year 
and 3-year maps. The noise-to-signal ratio is normalized to 1 for 
1-year map from detector $w_1$ alone.}
\end{center}
\end{figure}

\begin{figure}[htp]
\begin{center}
\includegraphics[height=8cm,width=12cm]{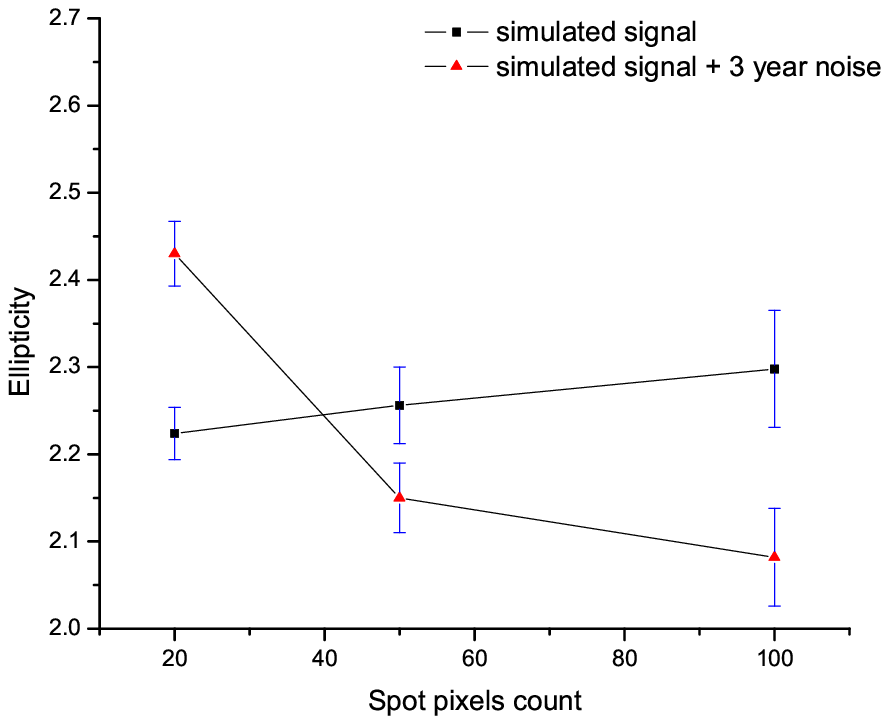}
\caption{Average ellipticity in 50 CMB maps simulated for the power spectrum 
parameters with and without noise. The inconsistency with Fig.6 is visible
already in the non-monotonic behavior of the noise for spots with
various pixel counts.}
\end{center}
\end{figure}


\bigskip


\begin{thebibliography}{99}
\bibitem{Bern1} de Bernardis P., et al,  2000, Nature, 404, 955.
\bibitem{Bern2} de Bernardis P., et al, 2003, IAU Symposium 216: Maps of the Cosmos, ASP Conf. Series;
astro-ph/0311396.
\bibitem{ellipse} Gurzadyan V.G., Ade P.A.R., de Bernardis P. et al, 2003, Int.J.Mod.Phys. D12, 1859.
\bibitem{Ben03} Bennett C.L., et al., 2003, ApJS 148, 1.
\bibitem{WMAP1year} Gurzadyan V.G., de Bernardis P. et al, 2005, Mod.Phys.Lett. A20, 893; 2003, Nuovo Cimento, B118,1101.
\bibitem{GT} Gurzadyan V.G., Torres S., 1997, A \& A, 321, 19.
\bibitem{Gurz1} Gurzadyan V.G., Ade P.A.R., de Bernardis P. et al, 2005, Mod.Phys.Lett., A20, 491.
\bibitem{Erik}
Eriksen H.K. et al, 2004, ApJ, 605, 14.
\bibitem{Copi1}
Copi C.J., Huterer D., Starkman G.D., 2004, Phys. Rev. 70, 043515
\bibitem{Sch}
Schwarz D.J., et al, 2004,  Phys.Rev.Lett. 93, 221301
\bibitem{Land}
Land K., Magueijo J., 2005, Phys. Rev. D72, 101302. 
\bibitem{Copi2}
Copi C.J., Huterer D., Schwarz D.J., Starkman G.D., 2006,  Mon.Not.Roy.Astron.Soc. 367, 79
\bibitem{Mag}
Magueijo J., Sorkin R.D., 2006, astro-ph/0604410
\bibitem{Cr}
Cruz M., Cayon L., Martinez-Gonzalez E., et al, 2006, astro-ph/0603859
\bibitem{Mc}
McEwen J. D., Hobson M.P., Lasenby A.N., Mortlock D.J., 2006, astro-ph/0604305 
\bibitem{Sp} Spergel D. et al, 2006, astro-ph/0603449
\bibitem{Page} Page L. et al, 2006, astro-ph/0603450
\bibitem{Hin} Hinshaw G. et al, 2006, astro-ph/0603451
\bibitem{GK1} Gurzadyan V.G., Kocharyan A.A., 1992, A \& A, 260, 14; Europhys.Lett. 1993, 22, 231.
\bibitem{Pen} Penrose R., {\it The Road to Reality}, Jonathan Cape, London, 2004.
\bibitem{Anosov} Anosov D.V., 1967, Comm. Steklov Mathematical Inst., vol.90.
\bibitem{G} Gurzadyan V.G., 1999, Europhys.Lett., 46, 114.
\bibitem{Bond} Bond J.R., Efstathiou G., 1987, MNRAS, 226, 655.
\bibitem{Efs03} Efstathiou G., 2003, MNRAS,  343, L95.
\bibitem{Lum03} Luminet J.-P. et al, 2003, Nature, 425, 593.
\bibitem{Aur04} Aurich R., Lustig S., Steiner F., Then H., 2005, Phys.Rev.Lett. 94, 021301.
\bibitem{Gorski} G\'orski, K.M., Hivon, E. and
 Wandelt, B.D., 1998, astro-ph/9812350; http://www.eso.org/kgorski/healpix/
\bibitem{Copi} Copi C.J., Huterer D., Schwarz D.J., Starkman G.D., 2006, astro-ph/0605135
\bibitem{DL} Draine B.T., Lazarian A. 1998, ApJ, 494, L19.
\bibitem{GX}
Gurzadyan V.G., Xue S.S. 2003, Mod.Phys.Lett. A18, 561.
\bibitem{V} 
Vereshchagin G.V., Yegorian G., 2006, Phys.Lett. B636, 150; Class. Quant.Grav. 23, 5049; astro-ph/0610197. 
\bibitem{DG}
Djorgovski S.G., Gurzadyan V.G., 2006, Nucl. Phys. B (in press); astro-ph/0610204.


\end{thebibliography}
\end{document}